\documentclass{article}
\usepackage{spconf,amsmath,graphicx,enumitem,amsfonts}
\usepackage{mathrsfs}
\usepackage{multirow, makecell}
\usepackage{booktabs}
\usepackage{numerica}
\usepackage[flushleft]{threeparttable}
\usepackage{enumitem}
\usepackage{xspace}
\usepackage[T1]{fontenc}
\usepackage[absolute]{textpos}
\usepackage{tikz}
\usepackage{mdframed}

\newcommand\thefont{\expandafter\string\the\font}

\AtBeginDocument{%
  \setlength\abovedisplayskip{0pt}
  \setlength\belowdisplayskip{0pt}}

\usepackage[draft,textsize=footnotesize,textwidth=15mm]{todonotes}
\newcommand{\modelname}{{NESSA}\xspace} 

\title{Post-training embedding alignment for decoupling enrollment and runtime speaker recognition models}

\name{Chenyang Gao$^{\star}$ \thanks{Chenyang Gao performed this work while interning at Amazon.} \quad Brecht Desplanques$^{\dagger}$ \quad Chelsea J.-T. Ju$^{\dagger}$ \quad Aman Chadha$^{\dagger}$ \quad Andreas Stolcke$^{\dagger}$ } 
\address{$^{\star}$Rutgers, The State University of New Jersey \\ $^{\dagger}$Amazon Alexa AI, U.S.A.}
\begin{document}
\ninept
\maketitle
\begin{abstract}
Automated speaker identification (SID) is a crucial step for the personalization of a wide range of speech-enabled services. Typical SID systems use a symmetric enrollment-verification framework with a single model to derive embeddings both offline for voice profiles extracted from enrollment utterances, and online from runtime utterances. Due to the distinct circumstances of enrollment and runtime, such as different computation and latency constraints, several applications would benefit from an asymmetric enrollment-verification framework that uses different models for enrollment and runtime embedding generation. To support this asymmetric SID where each of the two models can be updated independently, we propose using a lightweight neural network to map the embeddings from the two independent models to a shared speaker embedding space. Our results show that this approach significantly outperforms cosine scoring in a shared speaker logit space for models that were trained with a contrastive loss on large datasets with many speaker identities. This proposed Neural Embedding Speaker Space Alignment (\modelname) combined with an asymmetric update of only one of the models delivers at least 60\% of the performance gain achieved by updating both models in the standard symmetric SID approach.
\end{abstract}
\begin{tikzpicture}[remember picture,overlay]
\node[anchor=south,yshift=10pt] at (current page.south) {\fbox{\parbox{\dimexpr\textwidth-\fboxsep-\fboxrule\relax}{\footnotesize \textcopyright 2024 IEEE. Personal use of this material is permitted. Permission from IEEE must be obtained for all other uses, in any current or future media, including reprinting/republishing this material for advertising or promotional purposes, creating new collective works, for resale or redistribution to servers or lists, or reuse of any copyrighted component of this work in other works.}}};
\end{tikzpicture}
\vspace{-1.0em}
\section{Introduction}
\vspace{-0.5em}
\label{sec:intro}
Speaker Identification (SID) systems are developed to recognize speakers by comparing their distinctive vocal characteristics. Current online SID systems extract speaker embeddings in real-time fashion from the incoming audio streams and perform speaker identification by comparing these embeddings against existing voice profiles~\cite{x-vector, wan2018generalized, pelecanos2021dr}. The voice profiles are created by averaging the embeddings across the registered utterances for each speaker. These systems utilize the same speaker embedding extractor during both the enrollment and verification stage. In the remainder of this paper, we will refer to this approach as the standard \textit{symmetric enrollment-verification framework}. However, recent research~\cite{asymmetric2022} has shed light on the potential of using different SID models for generating embeddings in each stage. This approach is referred to as an \textit{asymmetric enrollment-verification framework}. It eliminates the need to use the same model during the distinct stages of enrollment and verification and it leads to many potential practical applications. The key idea is to use \textit{embedding space alignment} to reduce the mismatch between embedding spaces originating from different SID models to enable direct embedding comparison.

This alignment opens up a range of potential applications. For example, Li \textit{et al.}~\cite{asymmetric2022} proposed to use asymmetric SID involving a larger model for generating embeddings during enrollment and a smaller model for embedding extraction from the runtime audio streams. During enrollment, a computationally intensive and noncausal model can be used to extract high-quality voice profiles, while the runtime model should exhibit minimal latency and computational cost.
Another pertinent application involves an industry-specific challenge. To comprehensively validate SID model performance in the real world and to compare the impact of different SID models, extensive A/B~\cite{siroker2015b} or A/B/n tests become a fundamental part of the evaluation pipeline. However, each candidate model in the standard symmetric enrollment-verification framework will require an updated voice profile for a vast set of enrolled speakers.
This poses a scaling issue when multiple model candidates are tested in parallel. This standard A/B/n test setup will also result in computation wasted on the creation of new voice profiles when some model candidates are eventually not being used. To overcome these inefficiencies, speaker embedding space alignment enables us to utilize the readily available voice profiles and to make them compatible with the candidate models, instead of creating new voice profiles for each candidate model.
Moreover, SID model updates would potentially impact downstream applications that rely on the generated speaker embeddings to provide extra speaker identity context. Embedding alignment would provide a path to updating the SID models, without significantly impacting the downstream applications by feeding those dependent systems the embeddings that have been aligned back to the original speaker embedding space.

Prior work in the speaker verification domain utilized a shared speaker logit score space to combine embeddings from different models to create a high-performing system ensemble \cite{silnova2022analyzing,avdeeva2021stc,novoselov2023universal}. This alignment depends on utterance-based score vectors containing the speaker similarity score against every individual training speaker in a large shared dataset that was used to train every individual system in the ensemble with a softmax-based classification loss. Even though the speaker logit score vectors can be produced by different SID systems, these score vectors can be directly compared through cosine similarity scoring, as the training speaker set is identical across the systems. In certain cases, cosine scoring in the speaker logit space can outperform cosine scoring in the speaker embedding space. It has also been shown that system fusion in this logit space outperformed the more standard score fusion~\cite{avdeeva2021stc,novoselov2023universal}.
However, the effectiveness of this scoring method remains uncertain when the SID models are trained with training objectives other than the typical softmax-based classification loss~\cite{deng2019arcface}. Classification-based loss functions are typically avoided when the number of training speakers becomes unmanageable for the classification head; in those cases one typically relies instead on scalable contrastive loss variants~\cite{wen2022sphereface,wan2018generalized, pelecanos2021dr}.
In another related study \cite{asymmetric2022}, an auxiliary loss was introduced to align speaker embedding spaces for various models during the training process. In~\cite{wang2019knowledge}, researchers proposed to use knowledge distillation to transfer the knowledge from a teacher model to a student model. While these methods alter the speaker embedding spaces to be aligned, they limit the flexibility of developing a new runtime and/or enrollment speaker embedding extractor completely independent from each other, since the alignment happens during training of the embedding extractor.

To account for the diversity of possible SID models and to allow for the models to be trained independently, we propose a flexible and lightweight Neural Embedding Speaker Space Alignment (\modelname) backend to align the speaker embeddings between frozen enrollment and runtime embedding extractors.
In the context of datasets with a very large numbers of speakers, our results showed that
speaker-logit-based alignment did not yield satisfactory results in the asymmetric enrollment-verification framework when the models were trained with different training objectives, speaker sets, and model structures.
\modelname on the other hand performed significantly better and can in certain scenarios completely close the performance gap when compared to a more costly update of both models in the standard symmetric enrollment-verification framework.

\vspace{-1.0em}
\section{Efficient Embedding Space Alignment}
\vspace{-0.5em}
\label{sec:related_work_method}
\subsection{Problem statement}
\vspace{-0.5em}
Consider two independently trained SID models denoted as model \textbf{\textit{X}} and~\textbf{\textit{Y}}, and corresponding speaker embedding spaces $\mathbb{E}_X$ and $\mathbb{E}_Y$, respectively. The goal is to conduct asymmetric speaker verification given the enrollment embeddings in $\mathbb{E}_X$ and the runtime embeddings in $\mathbb{E}_Y$.
Since \textbf{\textit{X}} and \textbf{\textit{Y}} are trained with different configurations or model architectures, for the various reasons described in Section~\ref{sec:intro}, $\mathbb{E}_X$ and $\mathbb{E}_Y$ are mismatched. The immediate task is to develop a space alignment approach that enables performant scoring in the asymmetric framework, without degrading the performance compared to the single-model symmetric system with the worst-performing model and to close the performance gap compared to the symmetric approach with the best-performing model.

\vspace{-1.0em}
\subsection{Speaker-logit-based embedding space alignment}
\vspace{-0.5em}
\label{sec:speaker_logit}
Speaker verification in the speaker logit space involves cosine scoring between score vectors that express the speaker similarity of an utterance against every individual speaker within a predefined set of speakers. These speaker similarities are typically estimated on the speakers that were used to train the embedding extractor. Thus, the embeddings of different model versions can be made compatible by calculating a lightweight mapping between the different speaker embedding spaces to the speaker logit score vectors based on a shared pool of training speakers~\cite{silnova2022analyzing,avdeeva2021stc,novoselov2023universal}. When both systems are trained with a classification-based loss, the speaker logits $\mathbf{l}$ refer to the high-dimensional last layer output of the model that is used as input for the softmax-based classification loss. They are computed as $\mathbf{l}=\mathbf{W}\mathbf{r}$, where $\mathbf{r}$ is the speaker embedding, and $\mathbf{W}$ is the classification weight matrix with shape $(N \times d)$ that defines the classification head. $N$ and $d$ denote the number of speakers in the training set and the embedding dimension, respectively. The final classification layer does not typically include a bias term~\cite{deng2019arcface}.

However, models that are trained with other training criteria such as contrastive losses~\cite{wan2018generalized} or binary cross entropy~\cite{wen2022sphereface} do not have such a classification weight matrix. To enable speaker logit scoring in this case, we construct a classification weight matrix post-training using voice profiles: $\mathbf{W} = \lbrack \mathbf{e}_{1}; \mathbf{e}_{2};  \hdots; \mathbf{e}_{N} \rbrack^T$.
Voice profile $\mathbf{e}_{i}$ indicates the length-normalized average enrollment embedding for speaker $i$, and $N$ is the number of selected speakers to construct $\mathbf{W}$. We perform speaker verification using cosine similarity scoring~$s_c$ for different models in the speaker logit space as follows:
\begin{align*}
  s_c(\mathbf{l}_{e},\mathbf{l}_{r})&=\frac{\mathbf{l}_{e}^T\mathbf{l}_{r}}{||\mathbf{l}_{e}||\cdot||\mathbf{l}_{r}||}\\
  &=\frac{\mathbf{e}_{X}^T\mathbf{W}_{X}^T\mathbf{W}_{Y}\mathbf{r}_{Y}}{||\mathbf{W}_{X}\mathbf{e}_{X}||\cdot||\mathbf{W}_{Y}\mathbf{r}_{Y}||} \\
  &=\frac{\mathbf{e}_{X}^T\mathbf{W}_{X}^T\mathbf{W}_{Y}\mathbf{r}_{Y}}{\sqrt{\mathbf{e}_{X}^T\mathbf{W}_{X}^T\mathbf{W}_{X}\mathbf{e}_{X}} \cdot \sqrt{\mathbf{r}_{Y}^T\mathbf{W}_{Y}^T\mathbf{W}_{Y}\mathbf{r}_{Y}}}
\end{align*}
where $\mathbf{l}_{e}$ and $\mathbf{l}_{r}$ are the speaker logit score vectors for the enrollment profile and runtime embedding, respectively. Matrices $\mathbf{W}_{X}$, $\mathbf{W}_{Y}$ represent the classification weights using a shared set of speakers for models \textbf{\textit{X}} and \textbf{\textit{Y}}. Embedding $\mathbf{e}_{X}$ is the enrollment voice profile generated by model~\textbf{\textit{X}} and $\mathbf{r}_{Y}$ is the runtime speaker embedding extracted by model \textbf{\textit{Y}}.

To make this scoring approach more efficient, we used the Cholesky decomposition and the fusion approach as described in~\cite{silnova2022analyzing}:
\begin{align*}
  s_c(\mathbf{l}_{e}, \mathbf{l}_{r})&=\frac{\Tilde{\mathbf{e}}_{X}^T\Tilde{\mathbf{W}}^T\Tilde{\mathbf{W}}\Tilde{\mathbf{r}}_{Y}}{\sqrt{\Tilde{\mathbf{e}}_{X}^T\Tilde{\mathbf{W}}^T\Tilde{\mathbf{W}}\Tilde{\mathbf{e}}_{X}} \cdot \sqrt{\Tilde{\mathbf{r}}_{Y}^T\Tilde{\mathbf{W}}^T\Tilde{\mathbf{W}}\Tilde{\mathbf{r}}_{Y}}} \\
  &=\frac{\Tilde{\mathbf{e}}_{X}^T\Tilde{\mathbf{M}}^T\Tilde{\mathbf{M}}\Tilde{\mathbf{r}}_{Y}}{\sqrt{\Tilde{\mathbf{e}}_{X}^T\Tilde{\mathbf{M}}^T\Tilde{\mathbf{M}}\Tilde{\mathbf{e}}_{X}} \cdot \sqrt{\Tilde{\mathbf{r}}_{Y}^T\Tilde{\mathbf{M}}^T\Tilde{\mathbf{M}}\Tilde{\mathbf{r}}_{Y}}} \\
  &=s_c(\Tilde{\mathbf{M}}\Tilde{\mathbf{e}}_{X}, \Tilde{\mathbf{M}}\Tilde{\mathbf{r}}_{Y})
\end{align*}
where $\Tilde{\mathbf{W}} = [\mathbf{W}_{X}; \mathbf{W}_{Y}]$ with shape $N \times 2d$,  $\Tilde{\mathbf{e}}_{X}^T=[\mathbf{e}_{X}^T;\mathbf{0}]$, and $\Tilde{\mathbf{r}}_{Y}^T=[\mathbf{0};\mathbf{\mathbf{r}}_{Y}^T]$.
$\Tilde{\mathbf{M}}$ is an upper triangular matrix with dimensions $2d \times 2d$ such that $\Tilde{\mathbf{M}}^T\Tilde{\mathbf{M}}=\Tilde{\mathbf{W}}^T\Tilde{\mathbf{W}}$. This approach allows for efficient speaker logit scoring that does not negatively scale with the number of training speakers $N$ in $\mathbf{W}$ during inference. 
\vspace{-1.0em}
\subsection{Neural Embedding Speaker Space Alignment}
\vspace{-0.5em}
\label{sec:learning_based}
Instead of performing the speaker similarity scoring in a shared speaker logit space, we propose to use a Neural Embedding Speaker Space Alignment (\modelname) that employs a lightweight DNN $\mathscr{F}$ to enable accurate cosine similarity scoring in the asymmetric enrollment-verification framework. In contrast to the approach of jointly training both models in the asymmetric framework as proposed in~\cite{asymmetric2022}, our proposal involves computing this space alignment after the training of each individual model.
As before, we assume that the length-normalized enrollment voice profile $\mathbf{e}_{X}^i$ from speaker~$i$ is produced by model~\textbf{\textit{X}}, and the length normalized runtime embedding $\mathbf{r}_{Y}^j$ from speaker~$j$ is generated by model~\textbf{\textit{Y}}.
 
We explore three different approaches to train NESSA:\\
\textbf{Scoring in embedding space~\textit{X}} ($\mathscr{M}_1$): In this approach, we use the embedding space of model \textbf{\textit{X}} as a reference space, and train a space aligner $\mathscr{F}$ to map runtime embeddings $\mathbf{r}_Y$ to model space \textbf{\textit{X}}, so as to perform the verification in that embedding space. The training objective is:
\begin{equation}
    \label{eq:NESSA_F1}
    \mathscr{L}=\frac{1}{N}\sum_i^N \text{MSE}(\mathscr{F}(\mathbf{r}_{Y}^i), \mathbf{r}_X^i)
\end{equation}
where we use the mean squared error (MSE) as the training objective and $N$ is the number of embeddings in each minibatch. During evaluation, we will perform cosine scoring between voice profile $\mathbf{e}_{X}$ and runtime embedding $\mathscr{F}(\mathbf{r}_{Y})$ in embedding space \textbf{\textit{X}}.\\
\textbf{Scoring in embedding space~\textit{Y}} ($\mathscr{M}_2$): Alternatively, we can perform this mapping the other way around by mapping the enrollment embeddings from space \textbf{\textit{X}} to space \textbf{\textit{Y}}, using loss
\vspace{-1.0em}
\begin{equation}
\vspace{-1.0em}
    \label{eq:NESSA_F2}
    \mathscr{L}=\frac{1}{N}\sum_i^N \text{MSE}(\mathscr{F}(\mathbf{e}_{X}^{i}), \mathbf{e}_{Y}^{i}).
\end{equation}
We then perform verification between $\mathscr{F}(\mathbf{e}_{X})$ and $\mathbf{r}_{Y}$ in embedding space \textbf{\textit{Y}}. An advantage of this approach is that the mapping of enrollment embeddings can be performed offline, which would completely eliminate the impact of \modelname on runtime latency.\\
\textbf{Boosting NESSA with contrastive learning} ($\mathscr{M}_3$):
The original embedding spaces~\textbf{\textit{X}} or \textbf{\textit{Y}} might not be the most suited spaces to compare the embeddings between two very different SID models. To further increase the impact of \modelname, we propose to adapt both the enrollment and runtime embeddings simultaneously to a new embedding space with two DNNs $\mathscr{F}_1$ and $\mathscr{F}_2$ respectively. We introduce an additional contrastive loss term as in~\cite{asymmetric2022} to make this new embedding space suitable for speaker verification purposes. As the NESSA backend will typically be trained on smaller datasets compared to the training dataset of the embedding extractors, we will still anchor the new embedding space to the embedding space of the best-performing model (here assumed to be model \textbf{\textit{Y}}) using MSE loss terms for both enrollment and runtime embeddings similar to Eq.~(\ref{eq:NESSA_F2}). The final loss function is defined as follows:
\begin{align}
    \label{eq:combined_loss}
    \mathscr{L} =& -\alpha \frac{1}{N}\sum_i^{N} \log\frac{e^{w \cdot s_c(\mathscr{F}_1(\mathbf{e}_{X}^{i}), \mathscr{F}_2(\mathbf{r}_{Y}^{i}))}}{\sum_j^{N+M} e^{w \cdot s_c(\mathscr{F}_1(\mathbf{e}_{X}^{j}), \mathscr{F}_2(\mathbf{r}_{Y}^{i}))}} \notag \\
    & + \beta \frac{1}{N}\sum_i^N \text{MSE}(\mathscr{F}_1(\mathbf{e}_{X}^{i}), \mathbf{e}_{Y}^{i}) \notag \\
    & + \gamma \frac{1}{N}\sum_i^N \text{MSE}(\mathscr{F}_2(\mathbf{r}_{Y}^{i}), \mathbf{r}_{Y}^{i}) 
\end{align}
where $\alpha, \beta, \gamma$ are scalars to control the importance of the loss function terms and $w$ is a trainable parameter to rescale the range of cosine similarity $s_c$. We will set a relatively low value for $\gamma$ as the corresponding loss term acts as a regularization penalty and does not help with learning a proper alignment between embedding spaces. Previous studies~\cite{tsai2021selfsupervised} showed that increasing the number of negative samples in contrastive learning leads to more discriminative representations. We increase the original number of negative samples in the contrastive loss term (limited by the batch size $N$) by adding $M$ additional distinct voice profiles.

\vspace{-1.0em}
\section{Experimental setup}
\vspace{-0.5em}
\subsection{Enabling quick A/B tests without voice profile updates}
We will use A/B testing as a case study. We will assume that candidate model~\textbf{\textit{Y}} outperforms the reference model~\textbf{\textit{X}} during offline evaluation. We have the existing voice profiles generated by model~\textbf{\textit{X}} and we want to enable cosine scoring with runtime embeddings extracted by the better model~\textbf{\textit{Y}} against the existing voice profiles through embedding alignment, instead of updating the voice profiles.
\label{sec:exp_setup}
\vspace{-1.0em}
\subsection{Datasets for embedding alignment and SID evaluation}
\vspace{-0.5em}
Training and evaluation is conducted on de-identified voice assistant speech data with consent of the speakers.
To construct the training dataset for embedding space alignment, we apply an existing speaker recognition model to the data and build positive speaker/utterance pairs based on high speaker similarity scores. This process results in a dataset with 200K speakers including both enrollment and runtime utterances.
Within the training dataset, instances from 10\% of the speakers serve as validation data for model selection and hyperparameter tuning.
The dataset for evaluating the SID systems is constructed by first randomly sampling de-identified utterances. The sampled utterances, together with the enrollment data of speakers associated with the same group of speakers, are compared by multiple annotators to create the speaker labels. We only keep utterances with consistent annotation labels. To evaluate the generalization capability of the trained models there is no group overlap between the training datasets and the evaluation data, but the alignment training dataset and evaluation datasets are sampled from the same in-domain distribution.
\vspace{-1.0em}
\subsection{Asymmetric SID systems}
\vspace{-0.5em}
To assess the effectiveness of the space alignment methods for varying performance progress, we select four SID models to construct two main asymmetric SID systems. The SID systems employ a multi-layer LSTM architecture~\cite{hochreiter1997long,wan2018generalized} with projection layers. Each LSTM layer has 1200 nodes, and 400 nodes in the projection layer. The output speaker embedding size is 400. The acoustic input features are 40-dimensional log Mel-filter bank energies with a Hamming window of 25\,ms and a step size of 10\,ms for all models. These features are passed through an energy-based voice activity detection module to remove the non-speech frames. The four models are:
\setlist{nolistsep}
\begin{itemize}[noitemsep,topsep=0pt]
    \item \textbf{$\text{GE2E}$}: A 3-layer LSTM architecture trained using the generalized-end-to-end (GE2E) loss in the default configuration from~\cite{wan2018generalized}. It was trained on a large internal voice assistant dataset, that is significantly larger than the embedding space alignment training datasets.
    \item \textbf{$\text{BCE}$}: A 4-layer LSTM architecture trained with the binary cross-entropy (BCE) loss~\cite{wen2022sphereface} on a second large internal dataset of the same scale as used for the \textbf{$\text{GE2E}$} model.
    \item \textbf{$\text{SA}_\textit{early}$}: A model that uses a 3-layer LSTM architecture trained with the GE2E loss on the space alignment (SA) training dataset with early stopping.
    \item \textbf{$\text{SA}_\textit{full}$}: Similar to \textbf{$\text{SA}_\textit{early}$} but trained until full convergence and initialized with a different random seed.
\end{itemize}
We define two asymmetric SID systems, each uniquely defined by their enrollment-verification model versions:
\setlist{nolistsep}
\begin{itemize}[noitemsep,topsep=0pt]
    \item \textbf{$\text{GE2E}$/$\text{BCE}$} enrollment-verification: The voice profiles are extracted by the \textbf{$\text{GE2E}$} model, while the runtime embeddings are generated by the \textbf{$\text{BCE}$} model. The goal is to evaluate embedding space alignment when both models have similar performance.
    \item \textbf{$\text{SA}_\textit{early}$/$\text{SA}_\textit{full}$} enrollment-verification: The voice profiles are extracted by the \textbf{$\text{SA}_\textit{early}$} model, while the runtime embeddings are generated by the \textbf{$\text{SA}_\textit{full}$} model. The main goal is to evaluate embedding space alignment when there is a large performance gap between the two models.
\end{itemize}

\vspace{-1.0em}
\subsection{Embedding space alignment configuration}
\vspace{-0.5em}
The speaker logit alignment weight matrix $\mathbf{W}$ is constructed from voice profiles generated by the enrollment embedding extractor for a varying number of speakers in the alignment training dataset. For example, $\Tilde{\mathbf{M}}_{1\rm K}$ indicates we are using $1000$ voice profiles to construct $\Tilde{\mathbf{W}}$ before executing the Cholesky decomposition.\\
The lightweight model architecture of \modelname is a 3-layer multi-layer perceptron (MLP) with ReLU activations~\cite{agarap2018deep}; the hidden size of the MLP is set to 800. The output embeddings are 400-dimensional.
Each model is trained for 50 epochs with 2000 training steps per epoch; the batch size is set to 1024. We used the Adam~\cite{adam2015} optimizer with an initial learning rate of $10^{-3}$, with an exponential learning rate decay with a ratio of 0.96 after every epoch. The weights in the loss function for \modelname with contrastive learning ($\mathscr{M}_3$) are set to $\alpha=1.0, \beta=0.5, \gamma=0.1$, $w$ is initialized to 5.

\begin{table*}[!ht]
\resizebox{2\columnwidth}{!}{
\begin{threeparttable}
\caption{Relative False Reject Rate (FRR) impact in \% of symmetric and asymmetric speaker verification at different fixed False Accept Rate (FAR) target values on an in-house evaluation dataset following the evaluation protocol described in~\cite{li21q_interspeech}. Higher relative FRR impact is better and 0\% impact indicates the baseline single-model symmetric systems.}
\label{table:results}
\centering

\begin{tabular}{ccccccccc}
\toprule
  &  & \multicolumn{3}{c}{\textbf{Relative FRR impact @ target FAR (\%)}}($\uparrow$) & & \multicolumn{3}{c}{\textbf{Relative FRR impact @ target FAR (\%) }}($\uparrow$) \\ 
 \multirow{2}{*}{\shortstack{\textbf{Embedding} \\ \textbf{Alignment Approach}}} &
 \multirow{2}{*}{\shortstack{\textbf{Enrollment/Verification} \\ \textbf{Model}}} &
 \multirow{2}{*}{\textbf{@12.5\%FAR}} &
 \multirow{2}{*}{\textbf{@5.0\%FAR}} &
 \multirow{2}{*}{\textbf{@2.0\%FAR}} &
 \multirow{2}{*}{\shortstack{\textbf{Enrollment/Verification} \\ \textbf{Model}}} &
 \multirow{2}{*}{\textbf{@12.5\%FAR}} &
 \multirow{2}{*}{\textbf{@5.0\%FAR}} &
 \multirow{2}{*}{\textbf{@2.0\%FAR}}
 \\ 
 \\
 \midrule
  $\times$ & $\text{GE2E}$/$\text{GE2E}$ & $0$ & $0$ & $0$ & $\text{SA}_\textit{early}$/$\text{SA}_\textit{early}$ & $0$ & $0$ & $0$ \\ 
 $\times$ & $\text{BCE}$/$\text{BCE}$ & $11.08$ & $8.55$ & $5.35$  & $\text{SA}_\textit{full}$/$\text{SA}_\textit{full}$ & $63.62$ & $62.67$ & $58.75$\\ 
 speaker logits $\Tilde{\mathbf{M}}_{\rm 200K}$ & $\text{GE2E}$/$\text{GE2E}$ & $-198.92$ & $-190.56$ & $-198.15$ & $\text{SA}_\textit{early}$/$\text{SA}_\textit{early}$ & $-36.75$ & $-20.44$ & $-13.65$ \\ 
 speaker logits $\Tilde{\mathbf{M}}_{\rm 200K}$ & $\text{BCE}$/$\text{BCE}$ & $-163.24$ & $-194.99$ & $-198.15$ & $\text{SA}_\textit{full}$/$\text{SA}_\textit{full}$ & $21.00$ & $26.19$ & $25.05$ \\ 
 \midrule
 speaker logits $\Tilde{\mathbf{M}}_{\rm 1K}$ &  $\text{GE2E}$/$\text{BCE}$ & $-514.59$ & $-497.20$ & $-464.79$ & $\text{SA}_\textit{early}$/$\text{SA}_\textit{full}$ & $-69.31$ & $-70.51$ & $-73.15$\\ 
 speaker logits $\Tilde{\mathbf{M}}_{\rm 10K}$ &  $\text{GE2E}$/$\text{BCE}$ & $-517.03$ & $-496.61$ & $-458.56$ & $\text{SA}_\textit{early}$/$\text{SA}_\textit{full}$ & $-62.88$ & $-65.79$ & $-71.22$ \\ 
 speaker logits $\Tilde{\mathbf{M}}_{\rm 200K}$ & $\text{GE2E}$/$\text{BCE}$ & $-504.86$ & $-488.94$ & $-461.19$ & $\text{SA}_\textit{early}$/$\text{SA}_\textit{full}$ & $-64.19$ & $-67.03$ & $-72.38$ \\ 
 \midrule
 NESSA $\mathscr{M}_1$ & $\text{GE2E}$/$\text{BCE}$ & $-1.35$ & $-2.36$ & $-3.50$ & $\text{SA}_\textit{early}$/$\text{SA}_\textit{full}$ & $6.94$ & $4.26$ & $4.62$ \\ 
 NESSA $\mathscr{M}_2$ &  $\text{GE2E}$/$\text{BCE}$ & $5.95$ & $4.13$ & $1.46$ & $\text{SA}_\textit{early}$/$\text{SA}_\textit{full}$ & $37.56$ & $36.12$ & $32.20$ \\ 
  NESSA $\mathscr{M}_3$ ($M=50\rm K$) & $\text{GE2E}$/$\text{BCE}$ & $11.35$ & $11.50$ & $7.30$ & $\text{SA}_\textit{early}$/$\text{SA}_\textit{full}$ & $43.62$ & $40.88$ & $35.48$ \\ 
\bottomrule
\end{tabular}
\vspace{-1.0em}
\end{threeparttable}
}
\end{table*}

\vspace{-1.0em}
\section{Results and analysis} 
\vspace{-0.5em}
\label{sec:results_and_analysis}
\subsection{Baseline results for symmetric enrollment-verification}
\vspace{-0.5em}
Baseline experiments involving a symmetric enrollment-verification framework are shown in the top rows of Table~\ref{table:results}. For all experiments we will report the relative false reject rate (FRR) changes at fixed target values of the false accept rate (FAR)~\cite{li21q_interspeech} against \textbf{$\text{GE2E}$} and \textbf{$\text{SA}_\textit{early}$} baselines. As expected asymmetric enrollment-verification without embedding space alignment did not perform significantly better than random scoring due to the mismatch of the embedding spaces, hence these results are not included.

\vspace{-1.0em}
\subsection{Speaker-logit-based embedding space alignment}
\vspace{-0.5em}
We present the speaker-logit-based embedding space alignment in the middle section of Table~\ref{table:results}. Speaker logit alignment enhances the results of the asymmetric framework compared to having no alignment at all. However, a six-fold FRR increase (around -500\%) is observed against a strong \textbf{$\text{GE2E}$} baseline. Additionally, increasing the number of alignment speakers only improves the performance marginally.
Table~\ref{table:results} also includes symmetric \textbf{$\text{GE2E}$/$\text{GE2E}$} and \textbf{$\text{SA}_\textit{early}$/$\text{SA}_\textit{early}$} speaker logit scoring. We observe that symmetric speaker-logit scoring triples the FRR (around -200\%)  when compared to the \textbf{$\text{GE2E}$} baseline that uses standard speaker embedding scoring which somewhat contradicts previous studies~\cite{avdeeva2021stc,silnova2022analyzing}. The degradation for symmetric speaker logit scoring with \textbf{$\text{SA}_\textit{early}$} is less pronounced (-10\% to -35\%), indicating it is important that the embedding extractors are trained on the same set of speakers as those used for speaker logit scoring, which significantly limits the flexibility of the alignment method. Most likely this performance gap can be further decreased by using classification-based losses to train the embedding extractors as proposed in~\cite{avdeeva2021stc,silnova2022analyzing}, however these types of losses cannot be directly applied to datasets with a large number of speakers, due to scaling issues.

\vspace{-1.0em}
\subsection{Neural Embedding Speaker Space Alignment}
\vspace{-0.5em}
The results with \modelname are presented in the bottom part of Table~\ref{table:results}. We observe the following.
First, all \modelname approaches perform significantly better than the speaker logit scoring method, demonstrating the effectiveness of training a post-training space embedding aligner using neural network techniques.
Second, $\mathscr{M}_2$ enrollment embedding alignment to the model candidate embedding space leads to significantly better results than $\mathscr{M}_1$  alignment to the original runtime embedding space. This is somewhat expected as the candidate model~\textbf{\textit{Y}} has better speaker verification performance, which should correspond to a higher-quality speaker embedding space; it should be the preferred target space for alignment. The performance of $\mathscr{M}_2$ is in between the performance of symmetric \textbf{$\text{GE2E}$} and \textbf{$\text{BCE}$}, showing that asymmetric framework with space alignment can benefit from updating a model on only a single side of the speaker verification trial.
Third, alignment $\mathscr{M}_3$ outperforms all other alignment methods. When the baseline and candidate model performance are comparable, as is the case for the \textbf{$\text{GE2E}$} and \textbf{$\text{BCE}$} models, $\mathscr{M}_3$ alignment can even slightly outperform the \textbf{$\text{BCE}$} candidate model in the symmetric framework. We argue this is because the alignment training data and the evaluation data are sampled from the same specific domain, and thus embedding alignment can perform (partial) finetuning. When the performance difference between the baseline and candidate models is large and the embedding extractors are trained on the same in-domain data  (\textbf{$\text{SA}_\textit{early}$} vs.~\textbf{$\text{SA}_\textit{full}$}), $\mathscr{M}_3$ can achieve up to 60\% of the performance improvement achieved by the candidate model in the symmetric framework.
\begin{table}[!t]
\vspace{-0.5em}
\resizebox{1.0\columnwidth}{!}{
\begin{threeparttable}
\caption{Ablation study for $\mathscr{M}_3$, on $\alpha$, $\beta$ and $\gamma$ and number of additional speakers $M$}
\label{table:ablation_study}
\centering
\begin{tabular}{ccccc}
\toprule
\textbf{Alignment Approach} & \textbf{Enrollment/Verification}  & 12.5\% & 5.0\% & 2.0\% \\
\midrule
$\times$ & $\text{GE2E}$/$\text{GE2E}$ & $0$ & $0$ & $0$ \\
$\mathscr{M}_3$ ($M=50\rm K$) & $\text{GE2E}$/$\text{BCE}$ & $11.35$ & $11.50$ & $7.30$ \\
\midrule
$\mathscr{M}_3$ ($\alpha=0$) & $\text{GE2E}$/$\text{BCE}$ & $7.57$ & $4.57$ & $-2.82$ \\
$\mathscr{M}_3$ ($\beta=0$, $\gamma=0$) & $\text{GE2E}$/$\text{BCE}$ & $-65.14$ & $-41.00$ & $-26.46$ \\
$\mathscr{M}_3$ ($M=0$) & $\text{GE2E}$/$\text{BCE}$ & $12.16$ & $6.93$ & $4.18$ \\
$\mathscr{M}_3$ ($M=10\rm K$) & $\text{GE2E}$/$\text{BCE}$ & $13.24$ & $10.77$ & $8.56$ \\
\midrule
$\times$ & $\text{SA}_\textit{early}$/$\text{SA}_\textit{early}$ & $0$ & $0$ & $0$ \\
$\mathscr{M}_3$ ($\beta=0$, $\gamma=0$) & $\text{SA}_\textit{early}$/$\text{SA}_\textit{full}$ & $14.06$ & $26.12$ & $28.12$ \\
\bottomrule 
\end{tabular}
\vspace{-1.5em}
\end{threeparttable}
}
\end{table}

Finally, we present an ablation study of $\mathscr{M}_3$ in Table~\ref{table:ablation_study} with the following findings. First, when excluding the effect of the contrastive loss by setting $\alpha=0$, the performance can already slightly improve over $\mathscr{M}_2$. This illustrates the benefit of realigning both spaces. Second, training an entirely new space by setting $\beta=0$ and $\gamma=0$ resulted in significantly worse performance. This highlights the importance of selecting a strong reference embedding space. We hypothesize that this caused by the fact that \textbf{$\text{GE2E}$} and \textbf{$\text{BCE}$} were already trained on a larger-scale dataset with only the SID task in mind. The construction of the new shared space is based on a smaller alignment dataset, which is detrimental for final SID performance. However, when there are significant performance differences between models due to a weaker \textbf{$\text{SA}_\textit{early}$} model as in the last row of Table~\ref{table:ablation_study}, the construction of a new space can perform better than the symmetric baseline. But the performance is still worse compared to utilizing a reference space in $\mathscr{M}_3$  or $\mathscr{M}_2$ alignment.

\vspace{-1.0em}
\section{Conclusion}
\vspace{-0.5em}
\label{sec:conclusion}
We have investigated post-training speaker embedding space alignment for SID systems within an asymmetric enrollment-verification framework, where different models are used to generate voice profiles and runtime speaker embeddings. A case study in enabling A/B tests within this asymmetric framework, so as to avoid extensive voice profiles rebuilding for each new candidate model, showed a need for embedding alignment. Our proposed \modelname method effectively bridges the mismatch between different embedding spaces, so that between 60\% and 100\% of the potential gain from the candidate model is achievable without explicit voice profile updates.



\vfill
\pagebreak


\bibliographystyle{IEEEbib}
\bibliography{refs}
\end{document}